\documentclass[11pt,prb,amsmath,amssymb]{revtex4}
\usepackage{latexsym}
\usepackage{amsfonts,amsbsy}

\def\onehalf{\textstyle{\frac{1}{2}}}

\def\onefourth{\textstyle{\frac{1}{4}}}

\def\gammabol{{\stackrel{\circ}{\Gamma}}{}}
\def\omegabol{{\stackrel{\circ}{\omega}}{}}

\def\Rbol{{\stackrel{\circ}{R}}{}}

\def\abol{{\stackrel{\circ}{a}}{}}

\def\Tbol{{\stackrel{\circ}{T}}{}}

\def\nablabol{{\stackrel{\circ}{\nabla}}{}}

\begin{document}

\title{SPIN AND ANHOLONOMY IN GENERAL RELATIVITY}

\author{R. Aldrovandi, P. B. Barros and J. G. Pereira}
\affiliation{Instituto de F\'{\i}sica Te\'orica,
Universidade Estadual Paulista\\
Rua Pamplona 145,
01405-900 S\~ao Paulo, Brazil}

\begin{abstract}
In the general case, torsion couples to the spin current of
the Dirac field. In General Relativity, the apparent torsion field to which the spin
current of the Dirac field couples is a mere manifestation of the tetrad anholonomy.
Seen from the tetrad frame itself, it has for components the anholonomy coefficients.
The latter represent mechanical characteristics of the frame. In the teleparallel
equivalent of General Relativity, this coefficient turns out to be the only torsion
present.
\end{abstract}

\maketitle
\renewcommand{\thesection}{\arabic{section}}
%%%%%%%%%%%%%%%%%%%%%%
\section{Introduction}
%%%%%%%%%%%%%%%%%%%%%%

Let a spacetime metric\cite{notation} $g_{\mu \nu}$ be given in terms of a tetrad
field  $h^a{}_{\mu}$ by
\begin{equation}
g_{\mu \nu} = \eta_{a b}\ h^{a}{}_{\mu} h^{b}{}_{\nu},
    \label{eq:tettomet}
\end{equation}
where $\eta_{a b}$ represents the Lorentz metric.
The metric determines a special connection, the Levi-Civita connection, whose
components are the Christoffel symbols
\begin{equation} %
\gammabol{}^{\lambda}{}_{\mu \nu} = {\textstyle
\frac{1}{2}} g^{\lambda \rho} \left[\partial_{\mu} g_{\rho \nu} +
\partial_{\nu} g_{\rho \mu} - \partial_{\rho} g_{\mu \nu} \right].
\label{Christoffel}
\end{equation}
Using Eq.(\ref{eq:tettomet}), it can be rewritten in terms of the tetrad field as
\begin{equation}
\gammabol{}^{\lambda}{}_{\mu \nu} = {\textstyle \frac{1}{2}}
\left\{ h_b{}^{\lambda} ( \partial_{\nu} h^{b}{}_{\mu} + \partial_{\mu}
h^{b}{}_{\nu}) + h^{a \lambda} h_a{}^\rho \left[ h_{a \nu}
(\partial_{\mu} h^{a}{}_{\rho} - \partial_{\rho} h^{a}{}_{\mu})
+ h_{a \mu} (\partial_{\nu} h^{a}{}_{\rho} - \partial_{\rho}
h^{a}{}_{\nu}) \right] \right\}.
\label{gammainh}
\end{equation}

Differently from the torsionless Levi-Civita connection, a general connection
$\Gamma^{\lambda}{}_{\nu \mu}$ exhibits torsion,\cite{KN96,AP95b} a tensor whose
components are essentially the antisymmetric parts of the connection components in
the last two indices:
\begin{equation}
    T^{\lambda}{}_{\mu \nu} = \Gamma^{\lambda}{}_{\nu \mu} -
    \Gamma^{\lambda}{}_{\mu \nu}.
    \label{eq:torsion}
\end{equation}
As said, this is zero for the Levi-Civita connection, but the existence of torsion
has important consequences even in that case: the property
$\Tbol^{\lambda}{}_{\mu \nu} = 0$ is at the origin of the well-known cyclic symmetry
of the Riemann tensor components.

An object is parallel-transported when its covariant derivative vanishes. 
The condition of metric compatibility (or preservation) is satisfied when the
metric is everywhere parallel-transported by the connection, that is,
\begin{equation}
\nabla_\lambda\ g_{\mu \nu} \equiv \partial_\lambda g_{\mu \nu} -
\Gamma^\rho{}_{\mu \lambda} g_{\rho \nu} - \Gamma^\rho{}_{ \nu
\lambda} g_{\mu \rho} = 0. \label{eq:compatibility0}
\end{equation}
We shall see later that any metric-preserving linear connection is, actually, a
Lorentz connection.
It is worth mentioning that a theorem by Ricci states that there is a single
connection with a given torsion and satisfying (\ref{eq:compatibility0}). The
Levi-Civita connection is that unique connection which is metric-compatible and has
zero torsion. If a connection $\Gamma$ preserves a metric and is not its Levi-Civita
connection, then it will have $T^{\lambda}{}_{\mu \nu} \ne 0$.

Tetrads represent frames.\cite{SC92,LL89} The components of general tensors, seen
from those frames, are obtained by contraction with $h^a{}_{\mu}$, and their inverses
$h_{b}{}^{\nu}$ are obtained from the relations 
\begin{equation}
h^{a}{}_{\mu} h_{a}{}^{\nu} = \delta_{\mu}^{\nu}\ \ {\rm and} \ \
h^{a}{}_{\mu} h_{b}{}^{\mu} = \delta^{a}_{b}. \label{eq:tetradprops1}
\end{equation}
Thus, a second rank tensor $V^{\lambda}{}_{\mu}$, for example, will be
seen with components
\[
V^{a}{}_{b} =
h^{a}{}_{\lambda} h^{\mu}{}_{b} \;
V^{\lambda}{}_{\mu}.
\]
Connections, however, follow a different rule.  The last index in
$\gammabol{}^{\lambda}{}_{\mu \nu}$ is a normal vector index, but the first two
have a different character. Seen from the tetrad frame, the Levi-Civita connection
appears as the spin-connection $\omegabol^{a}{}_{b c}$, related to the Christoffel
symbols by
\begin{equation} %
   \omegabol^{a}{}_{b c} \equiv \omegabol^{a}{}_{b \nu} h_c{}^\nu = h^{a}{}_{\lambda}\
    \gammabol^{\lambda}{}_{\mu \nu}\ h_{b}{}^{\mu} h_c{}^\nu +
    h_c{}^\nu h^{a}{}_{\rho} \ \partial_{\nu} \ h_{b}{}^{\rho}. 
    \label{eq:gamtomegabol} 
\end{equation} %
The essential role of a connection is to define covariant derivatives, such as 
${\nablabol}{}_\nu V^\lambda$ = $\partial_\nu V^\lambda + \gammabol^{\lambda}{}_{\mu
\nu} V^\mu$ for a vector field, and $\nablabol_\nu V_\lambda$ = $\partial_\nu
V_\lambda - \gammabol^{\mu}{}_{\lambda  \nu} V_\mu$ for a covector. In contrast with
the ordinary derivative, the covariant derivative of a tensorial object is another
tensorial object, and rule (\ref{eq:gamtomegabol}) ensures  its good behavior under
change to the tetrad frame:
\[
\nablabol_\nu V^\lambda = h^{a}{}_{\nu} h_b{}^\lambda \nablabol_a V^b.
\]

The name ``spin-connection'' given to (\ref{eq:gamtomegabol}) comes from its
appearence in the Dirac equation on a  Riemannian manifold,\cite{dirac}
\begin{equation} %
i \hbar \gamma^c h_c{}^\mu \left( \partial_\mu - {\textstyle{\frac{i}{4}}} \,
\omegabol{}^{a b}{}_{\mu} \, \sigma_{a b} \right) \psi \equiv i \hbar
\gamma^c \left( h_c - {\textstyle{\frac{i}{4}}} \, \omegabol{}^{a b}{}_{c} \,
\sigma_{a b} \right) \psi = m c \psi.
\label{cde}
\end{equation} 
This equation comes from the lagrangian
\begin{equation}
{\mathcal L}_\psi = {\textstyle{\sqrt{-g}}}
\left\{ {\textstyle{\frac{i}{2}}}
\left[ \bar{\psi} \gamma^\mu \left(\partial_\mu \psi -
{\textstyle{\frac{i}{4}}} \omegabol{}^{a b}{}_{\mu} \sigma_{a b} \psi \right) 
- \left(\partial_\mu  \bar{\psi} + {\textstyle{\frac{i}{4}}}
\omegabol{}^{a b}{}_{\mu}  \bar{\psi} \sigma_{a b}\right) \gamma^\mu \psi
\right] - {\textstyle{\frac{m c}{ \hbar}}}  \bar{\psi} \psi \right\}.
\label{dlgra}
\end{equation}

Let us introduce the tensors
\begin{equation}
\Theta^a{}_\mu = {\textstyle{\frac{i}{2}}} \left[ {\bar \psi} \gamma^a (\partial_\mu
\psi) - (\partial_\mu {\bar \psi}) \gamma^a \psi \right]
\end{equation}
and
\begin{equation}
S^{\mu}{}_{ab}
= -\ {\textstyle{\frac{1}{4}}}\ {\bar \psi} \left[  \gamma^\mu \sigma_{ab}
 + \sigma_{ab}  \gamma^\mu  \right] \psi 
= -\ {\textstyle{\frac{1}{4}}}\ {\bar \psi} \left\{\gamma^\mu, \sigma_{ab} \right\}
\psi,
\label{eq:spindensity}
\end{equation}
with $\gamma^\mu = \gamma^c h_c{}^\mu$. When $\psi$ is a solution of the Dirac
equation, $\Theta^a{}_\mu$ will be the canonical energy-momentum density tensor, and
$S^{\mu}{}_{ab}$ will be the spin density tensor. They can, however, be considered
also ``off-shell'', simply as densities. In terms of these tensors, the Lagrangian
(\ref{dlgra}) takes on the form
\begin{equation}
{\textstyle{\frac{1}{ \sqrt{- g}}}}\ {\mathcal L}_\psi = h_a{}^\mu
\Theta^a{}_\mu - {\textstyle{\frac{1}{2}}}
\omegabol^{ab}{}_{\mu} S^{\mu}{}_{ab} - {\textstyle{\frac{m c}{ \hbar}}} {\bar
\psi} \psi.
\label{eq:LagDir}
\end{equation}
It can be seen from this expression that the energy-momentum couples to the tetrad
field, and the spin current couples to the spin-connection. The tensor $S^{cab} =
h^{c}{}_\mu \, S^{\mu ab}$ is clearly antisymmetric in the last two indices. Actually,
using the basic
$\gamma$-matrices property
\begin{equation}
\gamma^a \gamma^b = \eta^{ab} - i \sigma^{ab},
\label{gamgam}
\end{equation}
as well as its consequences
\begin{equation}
\sigma^{ab} = {\textstyle{\frac{i}{2}}} [\gamma^a, \gamma^b], \quad
\{\gamma^a, \gamma^b\} = 2 \eta^{ab}, \quad \{\gamma^c, \sigma^{ab} \} = -\ 
\{\gamma^b, \sigma^{ac}\},
\label{eq:identitygs}
\end{equation}
it follows that $S^{cab}$ is totally antisymmetric, or has cyclic symmetry:
$$%
S^{cab} = -\ S^{bac} = S^{bca}. 
$$
The Lagrangian can, consequently, be rewritten as
\begin{equation}
{\mathcal L} 
=  \sqrt{- g} \, \left[ h_a{}^\mu \Theta^a{}_\mu - {\textstyle{\frac{1}{2}}}
\omegabol_{a[bc]} S^{cab} - {\textstyle{\frac{m c}{ \hbar}}} {\bar \psi} \psi
\right].
\label{eq:LagDir2}
\end{equation}
The piece $\omegabol_{a[bc]}$ of the spin connection, antisymmetric in the last
two indices as in (\ref{eq:torsion}), has the look of a torsion, but there is no
torsion available: the only connection involved is the zero-torsion Levi-Civita
connection. It is our aim here to show that indeed no real torsion is involved
because $\omegabol_{a[bc]}$ is a pure property of the tetrad frame. In order to do
so, we shall be forced to make a kind of revision of what has been said above, in
particular to clarify the meaning of the many objects involved in the standard
notation used. 

%%%%%%%%%%%%%%%%%%%%%%%%%%%%
\section{The role of frames}
%%%%%%%%%%%%%%%%%%%%%%%%%%%%
\label{sec:frames}

A coordinate system $\{x^\mu\}$ defines local bases for vector fields, formed by the
set of gradients $\{\frac{\partial \ \ }{\partial x^\mu}\}$, and also  for covector
fields --- formed by the differentials $\{dx^\mu\}$.  These bases (or frames) are
dual, in the sense that $dx^\mu(\frac{\partial\ \ }{\partial x^\nu})$ =
$\delta^\mu_\nu$. A vector field will be a derivative $U = U^\mu \partial_\mu$, a
covector will be a 1-form $\omega = \omega_\mu dx^\mu$. Pairs $\{\partial_\mu,
dx^\nu\}$ of bases, directly related to a coordinate system, are said to be
holonomic, coordinate or natural. The  $g_{\mu \nu}$'s in Eq.(\ref{eq:tettomet})  and
the  Christoffel symbols are the  components of the  metric $g$  and the Levi-Civita
connection $\gammabol$ in base $\{dx^\mu\}$. The notation $\{h_{a}, h^{a}\}$ will be
used for a generic tetrad field.

Tetrads --- bases on 4-dimensional spacetime --- are not necessarily of this kind.
Any set $\{h_a\}$ of four linearly independent vector fields is a local base, or 
linear frame, and has its dual $\{h^a\}$ with $h^a (h_b) = \delta^a_b $. A holonomic
tetrad will always be of the form $\{h_a = \frac{\partial\ \ }{\partial y^a}, h^{a} =
d y^a\}$ for some set of functions $\{y^a\}$. The components of each covector $h^a$
= $h^a{}_\mu dx^\mu$ along $dx^\mu$ are obtained by applying it to
$\frac{\partial \ }{\partial x^\mu}$: $h^a{}_\mu = h^a(\frac{\partial \ }{\partial
x^\mu})$. The procedure can be inverted when $h^a$ are linearly independent,
and defines vector fields $h_a$ = $h_a{}^\mu \frac{\partial \ }{\partial x^\mu}$
which are not gradients. This means that, given the commutation table
\begin{equation}
    [h_{a}, h_{b}] = f^{c}{}_{a b}\ h_{c},
    \label{eq:comtable}
\end{equation}
there will be non-van\-ishing structure coefficients $f^{c}{}_{a b}$ for some
$a, b, c$. The frame $\{\frac{\partial\ \ }{\partial x^{\mu}}\}$ is holonomic
 because its members commute with each other. The components of vector
$h_a$ = $h_a{}^\mu \partial_\mu$ along $\partial_\mu$ comes from applying
$dx^\mu$: $h_a{}^\mu = dx^\mu (h_a)$. Duality of $\{h_{a}, h^b\}$ is then ensured
by Eqs.(\ref{eq:tetradprops1}).

A more intuitive view of the meaning of $f^{c}{}_{a b}$ can be obtained in terms of
Lie derivatives, which generalize usual partial derivatives. The partial derivative
$\frac{\partial \psi}{\partial x^\mu}$ of an object $\psi$ is the Lie derivative of
$\psi$ along the local integral curve of vector field $\frac{\partial \ }{\partial
x^\mu}$, the coordinate axis $x^\mu$. The Lie derivative $h_{a} \psi$ will be the
derivative of $\psi$  along the local integral curve of vector field $h_{a}$.
Derivatives of vector fields are given by the commutator: $[h_{a}, h_{b}]$ is the Lie
derivative of the vector field $ h_{b}$ along the local integral curve of vector
field $h_{a}$. Consequently,
$f^{c}{}_{a b}$ in (\ref{eq:comtable}) measures the components of that
``partial'' derivative along $h_{c}$.
 
Denoting by $u$ the curve parameters, the components of a velocity $U$ are given by
the holonomic form $dx^\mu$ applied to the time-evolution vector field $\frac{d\
}{du}$, that is,
\[
U^\mu = \frac{dx^\mu}{du} = dx^\mu \left(\frac{d\ }{du} \right).
\]
The velocity $U^\mu$ represents, consequently, the variation of the coordinate
$x^\mu$ in time $u$.  Seen from the tetrad $\{h_a\}$, $U$ has components 
\begin{equation}
U^a = h^a{}_\mu U^\mu = h^a{}_\mu dx^\mu
\left(\frac{d\ }{du} \right) = h^a \left(\frac{d\ }{du} \right).
\label{Ua}
\end{equation}
If $\{h_a\}$ is holonomic, then $h^a$ = $d y^a$ for some coordinates $\{y^a\}$,
and $U^a = \frac{d y^a}{d u}$, measuring the
variation of coordinate $y^a$ in time $u$.  If $\{h_a\}$ is
not holonomic, however, $U^a$ will
be the variation of {\it no} coordinate with time (the standard
non-relativistic example of anholonomic velocity is the
angular velocity of a rigid body in the general, non-planar case). 

Bases $\{h_{a}, h^b\}$ will be anholonomic in the generic case. In fact, a
gravitational field is present only when the tetrad field in  (\ref{eq:tettomet})
is anholonomic. As $\partial_{\nu} h^{b}{}_{\mu} =
\partial_{\mu} h^{b}{}_{\nu}$ for  a holonomic tetrad, expression
(\ref{gammainh}) reduces to 
\begin{equation} %
\gammabol{}^{\lambda}{}_{\mu \nu} =  h_b{}^{\lambda}\partial_{\mu} h^{b}{}_{\nu} 
\label{gammainholo}
\end{equation} %  
and the covariant representative of the gravitational field in General
Relativity, the curvature Riemann tensor
\begin{equation}
\Rbol^\lambda{}_{\rho \mu\nu} = \partial_\mu \gammabol^{\lambda}{}_{\rho \nu} 
- \partial_\nu \gammabol^{\lambda}{}_{\rho \mu} +
\gammabol^{\lambda}{}_{\sigma \mu} \gammabol^{\sigma}{}_{\rho \nu} -
\gammabol^{\lambda}{}_{\sigma \nu} \gammabol^{\sigma}{}_{\rho \mu},
\label{Rbol}
\end{equation}
vanishes. In that case, $g_{\mu \nu}$ would be simply the components of the
Lorentz metric $\eta$ transformed to the coordinate system $\{x^\mu\}$.

The dual version of the commutation table (\ref{eq:comtable}) is the Cartan structure
equation
\begin{equation}
    d h^{c} = -\ \onehalf f^{c}{}_{a b}\ h^{a} \wedge h^{b} =   \onehalf \ 
(\partial_\mu h^c{}_\nu - \partial_\nu h^c{}_\mu)\ dx^\mu \wedge dx^\nu.
    \label{eq:dualcomtable}
\end{equation}
The structure coefficients --- or anholonomy coefficients --- represent the
curls of the base members:
\begin{equation}
f^c{}_{a b}  = h_a{}^{\mu} h_b{}^{\nu} (\partial_\nu
h^c{}_{\mu} - 
\partial_\mu h^c{}_{\nu} ) =  h^c{}_{\mu} [h_a(h_b{}^{\mu}) - 
h_b(h_a{}^{\mu})]. \label{fcab}
\end{equation}
If $f^{c}{}_{a b}$ = $0$, then $d h^{a} = 0$ implies the local existence of functions
(coordinates) $y^a$ such that $h^{a}$ = $dy^a$. The tetrads are gradients when the
curls vanish.  

Combining expression (\ref{fcab}) with (\ref{eq:gamtomegabol}), the true meaning
of the ``torsion'' appearing in  (\ref{eq:LagDir2}) is found:
\begin{equation} %
	\omegabol^{a}{}_{b c} - \omegabol^{a}{}_{c b} =
f^{a}{}_{c b}.
\label{omandf}
\end{equation}
It follows that that Lagrangian is also
\begin{equation}
{\mathcal L} 
=  \sqrt{- g} \left\{ h_a{}^\mu \Theta^a{}_\mu + {\textstyle{\frac{1}{4}}}
f_{abc} S^{cab} - {\textstyle{\frac{m c}{ \hbar}}} {\bar \psi} \psi \right\}.
\label{eq:LagDirnew}
\end{equation}
The spin tensor couples actually to the anholonomy  of the tetrad field ---
if we wish, to the completely antysymmetrized anholonomy coefficient. The Dirac
equation derived from (\ref{eq:LagDirnew}) is
\begin{equation} %
 i \hbar
\gamma^c \left( h_c + {\textstyle{\frac{i}{8}}} \, f_{ab c} \,
\sigma^{a b} \right) \psi = m c \psi.
\label{cdewithf}
\end{equation} 

The antisymmetric part acquired by a  symmetric connection like
$\gammabol$ when looked at from the frame $\{h_{a}\}$ is a mere effect
of the  basis anholonomy. Let us make it clear with an example from electromagnetism.
Seen from frame $\{h_{a}\}$, the  electromagnetic field strength  $F_{\mu \nu}$ =
$\partial_{\mu} A_{\nu} - \partial_{\nu} A_{\mu}$  will have components
$$%
F_{ab} = h_{a}{}^{\mu} h_{b}{}^{\nu} F_{\mu \nu} = 
 h_{a} (A_{b}) - h_{b} (A_{a}) + (\omega^{c}{}_{ab} - \omega^{c}{}_{ba}) A_{c},
$$%
which is the same as
\begin{equation} %
F_{ab} = h_{a}
(A_{b}) - h_{b} (A_{a}) - f^{c}{}_{ab}  A_{c}.
\label{Finh2}
\end{equation} %
The last, anholonomy term is essential to the invariance of $F_{ab}$ under a $U(1)$
gauge transformation as seen from the frame $\{h_{a}\}$, which is
\[
A_{a} \rightarrow A'_{a} = A_{a} + h_{a} \phi.
\]
The corresponding field strength transformation is consequently
\[
F'_{ab} = h_{a} A'_{b} - h_{b} A'_{a} - f^{c}{}_{ab} 
A'_{c} = F_{ab} + h_{a} h_{b} \phi - h_{b} h_{a} \phi -
f^{c}{}_{ab} h_{c} \phi = F_{ab}.
\]
Now, it so happens that  (\ref{Finh2}) is exactly what comes out from a direct
calculation of the invariant form $F = dA$ = $d(A_{a} h^{a})$ by using
(\ref{eq:dualcomtable}) {\em in the absence of any Lorentz connection}.  

This is enough to vindicate our main claim, but it is interesting to learn more
about the spin connection and its relationship with anholonomy. The latter
actually determines $ \omegabol^a{}_{b c}$ completely. Indeed, direct substitution
of (\ref{gammainh}) into (\ref{eq:gamtomegabol}), and use of (\ref{fcab}), leads to a
general relation, from which (\ref{omandf}) follows automatically. This equation, in
turn, is equivalent to
\begin{equation}
 \omegabol^a{}_{b c} = - {\textstyle \frac{1}{2}} 
\left(f^{a}{}_{bc} + f_{bc}{}^a{} - f_{c}{}^a{}_b \right).
\label{eq:ominfs}
\end{equation}

Let us recall the force equation:\cite{LL89,HE73} In absence of external (that is,
non-gravitational) forces, a structureless particle follows a path given by the
geodesic equation
\begin{equation}
    \frac{\nablabol U^{\lambda}}{\nabla u} \equiv \frac{dU^{\lambda}}{du} +
\gammabol^{\lambda}{}_{\mu \nu} U^{\mu}
    U^{\nu} = 0.
    \label{eq:geodesic1}
\end{equation}
The path is a curve $\gamma(u)$ whose velocity field $U$ itself is
parallel-transported by $\gammabol$ along the curve.  The first term in the
middle expression is the kinematic acceleration of Special Relativity. The
second is the effect of gravitation seen as  an inertial acceleration.
 
In the presence of an external force the particle acquires an aceleration given by
the so-called equation of force
\begin{equation}
\abol^\lambda = \frac{\nablabol U^{\lambda}}{\nabla u}
\label{eq:force}
\end{equation}
which, seen from the tetrad frame $\{h_{a}\}$, takes the form
\begin{equation}
 \abol^a\  =   \frac{d U^{a}}{du} + \omegabol^{a}{}_{b c}\ U^{b}
U^{c}.
    \label{eq:force2}
\end{equation}
Use of (\ref{eq:ominfs}) allows to write the force equation  in terms of the
anholonomy coefficients:
\begin{equation}
   \frac{d U^{a}}{du} +
 f_{b}{}^a{}_c\ U^{b} U^{c} = \abol^a.
\end{equation} %
Here we see the second, inertial term for what it is: A frame effect. 

Expression (\ref{eq:ominfs}) shows a further
characteristic of the spin connection:
with the index lowered by the Lorentz metric, $\omegabol_{a b c}$
is antisymmetric in the first two indices: 
\begin{equation}
\omegabol_{abc} = -\ \omegabol_{bac}. \label{eq:omisLor}
\end{equation}
Of course, only $\omegabol_{[ab]c}$ appears in the Dirac equation, but it could
happen that the connection had some part $\omegabol_{(ab)c}$ which the equation
would ignore. We see that this is not the case. Many properties have been used to
arrive at (\ref{eq:ominfs}), but  one is enough to give (\ref{eq:omisLor}): it
comes directly from  (\ref{eq:compatibility0}) written for the Levi-Civita connection,
which  reads
\begin{equation}
	\partial_\lambda g_{\mu \nu} = 2\ \gammabol_{(\mu \nu) \lambda} =\gammabol_{\mu
\nu
\lambda} + \gammabol_{\nu \mu \lambda} = h^a{}_{\mu}h^b{}_{\nu }h^c{}_{\lambda}
(\omegabol_{abc} + \omegabol_{bac}) + \partial_\lambda g_{\mu \nu}.
\label{compatibility}
\end{equation}
We can alternatively say that, from the tetrad point of view,
(\ref{eq:compatibility0}) is
\begin{equation}
h_c (\eta_{ab}) - \omegabol^d{}_{ac} \eta_{db} - \omegabol^d{}_{bc} \eta_{ad} = 0.
\end{equation}
The underlying content of metric-preservation is revealed by property
(\ref{eq:omisLor}), which states that
the connection is Lorentzian. Let us see why.

%%%%%%%%%%%%%%%%%%%%%%%%%%%%%%%%%%%%%%%%
\section{Linear and Lorentz connections}
%%%%%%%%%%%%%%%%%%%%%%%%%%%%%%%%%%%%%%%%

Connections are 1-forms with values in Lie algebras of Lie groups. Gauge
potentials are related to groups not involving spacetime. For example, the gauge
potential of chromodynamics, the gluon field, is a connection  with values in Lie
algebra of the group $SU(3)$-color. The potential $A_\mu$ of electrodynamics does not
exhibit this character clearly because it has values in Lie algebra of the group
$U(1)$, which is 1-dimensional.

General Relativity works with spacetime itself, seen as a differentiable manifold.
Tensors and connections on a 4-dimensional  differentiable manifold are related to
the  real linear group $GL(4,\mathbb{R})$, the set of invertible $4 \times 4$
matrices with real entries. Connections with values in the Lie algebra
$gl(4,\mathbb{R})$ of $GL(4,\mathbb{R})$ are called linear connections. 

A  Lie algebra is a vector space with a binary internal operation which is
antisymmetric and satisfies the Jacobi identity. Matrices do constitute  Lie algebras
with the operation defined by the commutator. The algebra $gl(4,\mathbb{R})$ is
formed by all $4 \times 4$ matrices with real entries. In discussing such algebras,
it is extremely convenient to establish a base for the underlying vector space. The
simplest in the case is formed by the matrices $\Delta_{a}{}^{b}$ whose entries are
all zero except for that  of the $a$-th row and $b$-th column, which is equal to 1: 
\begin{equation} %
(\Delta_{a}{}^{b})^{d}{}_{c} = 
\delta^{d}_{a} \, \delta^{b}_{c}. 
\label{invDeltas}
\end{equation} %
An arbitrary $4 \times 4$ matrix $K$ can be written $K$ = 
$K^{a}{}_{b} \, \Delta_{a}{}^{b}$. The rather  
awkward index positions are chosen to make contact with the standard notation of
General Relativity.  
Linear connections are matrices of 1-forms, 
\begin{equation}
\Gamma = \Delta_{a}{}^{b} \Gamma^{a}{}_{b \mu} dx^{\mu} = \Delta_{a}{}^{b}
\Gamma^{a}{}_{b c} h^{c}.
\end{equation}

A general procedure allows to obtain subalgebras of $gl(4,\mathbb{R})$
related to the orthogonal or pseudo-orthogonal subgroups of $GL(4,\mathbb{R})$. We
shall apply it directly for the Lorentz group. Define first new matrices with indices
lowered by the Lorentz metric, $\{\Delta_{ab}\}$ = $\eta_{bc}\Delta_a{}^c$. Then,
the set of matrices $\{J_{a b} = \Delta_{ab} - \Delta_{ba}\}$ provide a (vector)
representation for the generators of the Lorentz group. Their entries are 
\begin{equation}
(J_{cd})^a{}_b =
\eta_{db}\ \delta^a_c - \eta_{cb}\ \delta^a_d. \label{Lorgens}
\end{equation}
Notice that we have simply lowered the indices with the pseudo-orthogonal  Lorentz
metric and antisymmetrized. Consequently, any linear connection which appears
antisymmetric in the first two indices and has them raised and lowered by the 
Lorentz metric is a connection with values in the Lie algebra of the  Lorentz group.
This is the case of the spin connection, and will be the case of any connection
preserving the metric (\ref{eq:tettomet}): Any such connection will have, once seen
from the tetrad frame, the form 
\begin{equation}
\omega = \onehalf\ J_a{}^b \, \omega^a{}_{bc} \, h^c,
\end{equation}
with $J_a{}^b$ the Lorentz generators. 

A Lorentz transformation will be given by 
\begin{equation}
\Lambda = 
\exp \left[{\onehalf \, J_{c d} \, \alpha^{c d}} \right],
\label{Lor1}
\end{equation}
where $\alpha^{c d}$ are the transformation parameters:  $\alpha^{ij}$ are the usual
rotation angles, and  $\alpha^{0k}$ are the boost parameters --- essentially relative
velocities. To see more about Lorentz connections it is necessary to look at the
place taken by the Lorentz group in General Relativity.
 
%%%%%%%%%%%%%%%%%%%%%%%%%%%%%%%%%%%
\section{Role of the Lorentz group}
%%%%%%%%%%%%%%%%%%%%%%%%%%%%%%%%%%%

It is no surprise that the Lorentz group somehow turns up. A spacetime is a
4-dimensional differential manifold whose tangent space is, at each point, a
Minkowski space on which it acts. A tetrad field  $h^a$ takes the Lorentz metric
$\eta_{a b}$ into a spacetime metric $g$, which has the components given in
Eq.(\ref{eq:tettomet}) in the base $\{dx^\mu\}$, in terms of which 
$h^a = h^{a}{}_{\mu} dx^\mu$. Each metric $g = g_{\mu \nu} dx^\mu dx^\nu$ will make
of spacetime a particular Riemannian space. 

Given a gravitational field, the base $\{h_{a}\}$ which determines the corresponding
Riemannian metric is far from being unique. At each point of the Riemannian space,
Eq.~(\ref{eq:tettomet}) only determines the tetrad field up to Lorentz
transformations in the tetrad indices. Suppose in effect another tetrad $\{h'_{a}\}$
such that
\begin{equation}
g_{\mu \nu} = \eta_{a b}\ 
h^{a}{}_{\mu} h^{b}{}_{\nu} = \eta_{c d}\ 
h^{' c}{}_{\mu} h^{' d}{}_{\nu}. \label{etatogmunu}
\end{equation}
Contracting both sides with $h_{e}{}^{\mu}h_{f}{}^{\nu}$, we arrive at
\[%
\eta_{a b} = \eta_{c d}\ 
(h^{' c}{}_{\mu} h_{a}{}^{\mu} ) (h^{' d}{}_{\nu} h_{b}{}^{\nu}).
\]%
This equation says that the matrix with entries 
\begin{equation}
\Lambda^a{}_{b} = h^{' a}{}_{\mu}\ h_{b}{}^{\mu},
\label{Lortetrad}
\end{equation}
which gives the transformation 
\begin{equation}
 h^{' a}{}_{\mu} = \Lambda^a{}_{b}\  h^{b}{}_{\mu},
\end{equation}
satisfies
\begin{equation}
 \eta_{cd} \ \Lambda^c{}_{a}\ \Lambda^d{}_{b} = \eta_{a b}.
\end{equation}
This is just the condition that a matrix $\Lambda$ must satisfy in
order to belong to (the vector representation of) the Lorentz group.

Besides the mentioned behavior when changing between holonomic and anholonomic bases, 
condition (\ref{eq:gamtomegabol}) ensures the covariant deri\-vative good behavior 
also under Lorentz transformations: $\nabla_\nu V^\lambda$ = $h^{a'}{}_{\nu}
h_{b'}{}^\lambda \nabla_{a'} V^{\prime b}$  and $\nabla_{a'} V^{b'}$ =
$\Lambda^c{}_{a'} \Lambda^{b'}{}_d \nabla_c V^{d}$.  

The Riemannian metric $g = (g_{\mu \nu})$ is a Lorentz invariant. In other words, 
any two tetrad fields as $\{h_{a}\}$ and $\{h'_{a}\}$ in (\ref{etatogmunu}) yields
the same $g$.  A metric corresponds to an equivalence class of tetrad fields, the
quotient of the set of all tetrads by the Lorentz group.  The sixteen fields
$h^{a}{}_{\mu}$ correspond, from the field-theoretical point of view, to ten degrees
of freedom --- like the metric --- once the equivalence under the six-parameter
Lorentz group is taken into account.

If we use $\Lambda^{a}{}_b$ as given by Eq.~(\ref{Lortetrad}), as well as its
inverse
\[
(\Lambda^{-1})^{a}{}_b = h^{a}{}_{\mu} \, h'_{b}{}^{\mu} = \eta_{bc}
\, \eta^{ad} \, \Lambda^{c}{}_{d} = \Lambda_b{}^a,
\]
we find how the components change under a Lorentz transformation:
\begin{equation}
\omega^{'a}{}_{b \nu} = \Lambda^a{}_c\ \omega^{c}{}_{d \nu}
(\Lambda^{-1})^{d}{}_b + \Lambda^a{}_c\ \partial_\nu
(\Lambda^{-1})^{c}{}_b.
\label{omisLor}
\end{equation}
This is just the usual behavior of a gauge potential under 
transformations of the corresponding gauge group. 

%%%%%%%%%%%%%%%%%%%%%%%%%%
\section{Inertial effects}
%%%%%%%%%%%%%%%%%%%%%%%%%%
\label{sec:nonin}

We have been talking about the effect of frames as inertial effects. Let us now see
that some of the anholonomy coefficients represent frame accelerations and angular
velocities.

The timelike member $h_{0}$ of a set $\{h_{a}\}$ of vector fields constituting a
tetrad will define, for each set of initial conditions, an integral curve
$\gamma$. It is always possible to identify $h_{0}$ to the velocity $U$ of $\gamma$.
This would mean $U^a = h^a{}_\nu h_0{}^\nu = \delta^a_0$. The frame, as it is carried
along that timelike curve, will be inertial or not, according to the corresponding
force law (\ref{eq:force}). The force equation can be obtained by using, for
example, Eq.~(\ref{eq:gamtomegabol}) written for
$h_{0}$:
\[%
\partial_{\nu} \, h_{0}{}^{\lambda} + \gammabol^{\lambda}{}_{\mu \nu} \,
h_{0}{}^{\mu} = h_{a}{}^{\lambda}\ \omegabol^{a}{}_{0 \nu}.
\]%
This leads, with $U = \frac{d\ }{du} = h_{0}$, to the expression
\begin{equation} %
h_{0}{}^{\nu} \partial_{\nu} \ h_{0}{}^{\lambda} +
\gammabol^{\lambda}{}_{\mu \nu}\ h_{0}{}^{\mu} h_{0}{}^{\nu} \equiv
U^\nu \partial_{\nu} \ U^{\lambda} + \gammabol^{\lambda}{}_{\mu \nu}\ U^{\mu}
U^{\nu}  = h_{a}{}^{\lambda}\
\omegabol^{a}{}_{0 \nu} h_{0}{}^{\nu},
\label{accel0}
\end{equation} %
implying the frame acceleration
\begin{equation} %
\abol{}^{\lambda}  = h_{k}{}^{\lambda} 
\omegabol^{k}{}_{0 0}  = h_{k}{}^{\lambda} f^0{}_{0 k}. 
    \label{accel4} 
\end{equation} %
It follows that an accelerated frame is necessarily anholonomic: it must
have {\em at least} $f^0{}_{0 k} \ne 0$.  

To examine the behavior of the spacelike members of the tetrad along $\gamma$,
consider  Eq.~(\ref{eq:gamtomegabol}) for $h_{i}$. Indicating by
$a_{(i)}{}^{\lambda}$ the covariant change rate of $h_{i}{}^{\lambda}$, 
\[
a_{(i)}{}^{\lambda} = \nabla_U h_{i}{}^{\lambda} =
h_{a}{}^{\lambda}\ \omegabol^{a}{}_{i 0}  = \onehalf h^{c \lambda}
(f_{ic0} + f_{0ci} + f_{c0i}).
\]
As
\[%
\nabla_U h_a{}^\lambda = h_c{}^\lambda\  \omegabol^c{}_{a \nu} U^\nu
\]%
for any $U$, the Fermi-Walker derivative\cite{MTW73} will be
\[%
\nabla_F h_a{}^\lambda = \nabla_U h_a{}^\lambda + a_a \,
U^\lambda - U_a \, a^\lambda.
\]%
The particular case
\[ %
\nabla_F h_0{}^\lambda = \nabla_U h_0{}^\lambda - U_0 \, a^\lambda = 0 
\] %
implies that $h_0$ is kept tangent along
the curve.  The other tetrad members, however, rotate with respect to a
Fermi-Walker observer with an angular velocity 
\begin{equation}
\omega^k = \onehalf\ \epsilon^{kij} \; \omegabol_{ij0} = -
\onefourth\ \epsilon^{kij} f_{0ij},
\label{Riccirot0}
\end{equation}
which shows $\omegabol^a{}_{b c}$ in their role of Ricci's coefficient of
rotation.\cite{SC92}

An ideal observer on spacetime in simply conceived in General Relativity
as a timelike curve with no self-intersections. Now, on such a curve there
exists  a tetrad field $\{H_a\}$ which is parallel-transported all
along.\cite{ABP02} That frame is inertial: by Eqs.(\ref{accel0}) and (\ref{accel4}), it
has vanishing acceleration. More than that, in that frame all components of $\omegabol$
vanish along the curve. The components seen from any other frame $\{h_a\}$ are given
by (\ref{omisLor}), 
\begin{equation}
\omegabol^{a}{}_{b \nu} = h^c{}_\nu\ \omegabol^{a}{}_{b c}=
(\Lambda^{-1})^a{}_c \partial_\nu \Lambda^{c}{}_b = (\partial_\nu \ln
\Lambda)^a{}_b.  \label{omegazerox}
\end{equation}
Using Eqs.(\ref{Lorgens}) and (\ref{Lor1}) it is found that
\begin{equation}
\omegabol^{a}{}_{b c} = h_c (\alpha^a{}_b).  \label{omegazerox2}
\end{equation}
The meaning of the above anholonomy coefficients emerge clearly:
Seen from frame  $\{h_a\}$ itself, its acceleration is
\begin{equation} %
f^{0k}{}_{0} = \abol{}^{k} =  
\omegabol^{k}{}_{0 0} = h_0 (\alpha^k{}_0),
\label{accel7} 
\end{equation} %
the time derivative of the boosts necessary to obtain it from an
inertial frame. Its angular velocity, in the same vein, is given by the 
time derivative of the rotation angles,
\begin{equation} f_{0ij} = -\ \omegabol_{ij0} = -\ \epsilon_{ijk}
\; \omega^k = -\ h_0 (\alpha_{ij}).
\end{equation}

%%%%%%%%%%%%%%%%%%%%%%%%
\section{Final comments}
%%%%%%%%%%%%%%%%%%%%%%%%

The usual statement of the Equivalence Principle is that, along any differentiable
non-selfintersecting worldline, it is possible to establish a {\em local} frame in
which the gravitational field is not felt. In a positive version of the Principle,
the gravitational field is locally replaced by an accelerated frame.\cite{CW95} The
spin current couples to the frame anholonomy, which contains both its linear
acceleration and its angular velocity. The spin term in the Dirac equation includes,
besides many other, the couplings of the electron velocity and spin to these
quantities.\cite{SP00} This is no novelty, except for one point: That acceleration
and that angular velocity are measured with respect to an inertial frame. It is usual
to describe them with respect to Fermi-Walker transported frames, of which inertial
frames are particular, but by far the most interesting physically cases.

A point that deserves to be mentioned is that, in the presence of torsion, Eq.
(\ref{omandf}) generalizes to
\begin{equation} %
\omega^{a}{}_{b c} - \omega^{a}{}_{c b} =
f^{a}{}_{c b} + T^{a}{}_{c b}.
\label{ometor}
\end{equation}
In the specific case of the teleparallel equivalent of General Relativity,\cite{tegr}
where $\omega^{a}{}_{b c} = 0$,\cite{hs} we see that the corresponding (Weitzenb\"ock)
torsion coincides with (minus) the anholonomy coefficient:
\begin{equation} %
T^{a}{}_{c b} = - f^{a}{}_{c b}.
\label{tptor}
\end{equation}
This result has interesting consequences. For example, in the same way the U(1) gauge
invariance of the Maxwell theory is not violated in a nonholonomous basis (see Section
\ref{sec:frames}), it is not violated either by the torsion of teleparallel
gravity.\cite{vector}

%%%%%%%%%%%%%%%%%%%%%%%
\begin{acknowledgments}
The authors thank Roberto Aureliano Salmeron for being who he is: a man and
a scientist worth emulating. They thank also FAPESP and CNPq for financial support.
\end{acknowledgments}

%%%%%%%%%%%%%%%%%%%%%%%%%%


\begin{thebibliography}{99}
\bibitem{notation}
The Latin alphabet $a, b, c, \dots = 0, 1, 2, 3$
will be used to denote Lorentz indices, which are raised and lowered with the Lorentz
metric $\eta$ = $\eta^{-1}$ = diag$(1,-1,-1,-1)$. Greek indices $\mu, \nu, \rho,
\dots = 0,1,2,3$ will refer to the Riemannian spacetime, and medium Latin indices $i,
j, k, \dots = 1, 2, 3$ to the corresponding space components.
\end{thebibliography}
\end{document}